\documentstyle[epsf,pra,aps,twocolumn]{revtex}

\draft

\begin{document}

\preprint{Submitted for publication in Physical Review A}

\title{A study of cross sections for excitation of pseudostates}

\author{Igor Bray\thanks{electronic address: I.Bray@flinders.edu.au}}
\address{
Electronic Structure of Materials Centre,
The Flinders University of South Australia,\\
G.P.O. Box 2100, Adelaide 5001, Australia}

\date{\today}
\maketitle

\begin{abstract}
Using the electron-hydrogen scattering Temkin-Poet model we
investigate the behavior of the cross sections for excitation of all
of the states used in the convergent close-coupling (CCC)
formalism. In the triplet channel, it is found that the cross section
for exciting the positive-energy states is approximately zero
near-threshold and remains so until a further energy, equal to the
energy of the state, is added to the system. This is consistent with
the step-function hypothesis [Bray, Phys. Rev. Lett. {\bf 78} 4721
(1997)] and inconsistent with the expectations of Bencze and Chandler
[Phys. Rev. A {\bf 59} 3129 (1999)]. Furthermore, we compare the
results of the CCC-calculated triplet and singlet single differential
cross sections with the recent benchmark results of Baertschy {\em et al.}
[Phys.  Rev. A (in press)], and find consistent
agreement.
\end{abstract}


In recent times the electron-impact ionization of atoms has attracted
considerable attention and controversy. Whereas we would argue that
the convergent close-coupling (CCC) approach to the problem
\cite{BF96} has been one of the most successful to date, Bencze and
Chandler~\cite{BC99} argue that there are fundamental flaws in the
method and that the presented results are inconsistent with the
prediction of their derivation in the limit of infinite CCC basis
sizes $N$. To be more precise, they first argue that their operator formalism
shows that in the limit of infinite $N$ the CCC formalism should
converge to the true ionization scattering amplitudes, which satisfy
the symmetrization postulate
\begin{equation}
f_S(\bbox{k},\bbox{q})=(-1)^Sf_S(\bbox{q},\bbox{k})
\label{symcon}
\end{equation}
in the case of e-H ionization, where $S=0,1$ is the total spin. 
Therefore, at any total (excess) energy
$E$ the resulting singly differential cross sections (SDCS) should be
symmetric about $E/2$. The fact that the CCC-calculated SDCS is not
symmetric Bencze 
and Chandler interpret as a lack of convergence of the calculations.

How can such a controversy be resolved? We are not able to perform
calculations with infinite basis sizes. Furthermore, we have already
acknowledged that there are convergence problems in the CCC
formalism~\cite{B97l}, but very different to those suggested by Bencze
and Chandler. Thus, the situation may appear to be somewhat
confused. The aim of this paper is to clarify the situation. We have
already given extensive arguments as to why we completely refute the
arguments of Bencze
and Chandler~\cite{B99reply}. Here we show that there are
unphysical consequences of their claims. Furthermore, to demonstrate
convergence, or lack of it, in the CCC method we compare with the
benchmark SDCS calculated by Baertschy {\em et al.}~\cite{BRIM99} using
the very recently developed external complex scaling (ECS) method. We 
shall only consider the Temkin-Poet model e-H problem~\cite{T62,P80}
due to its simplicity (only states of zero orbital angular momentum
involved), and yet it is sufficient to address all of the
issues involved. Accordingly, we shall write momenta as scalars.

In performing CCC calculations we first obtain a set of $N$ states
$\phi_n^{(N)}$ with energies $\epsilon_n^{(N)}$ ($n=1,\dots,N$) by
diagonalising the 
target Hamiltonian in an orthogonal Laguerre basis~\cite{BS92}. These
are then used to approximate the target space identity operator $I_2$
by
\begin{equation}
I_2\approx
I_2^{(N)}=\sum_{n=1}^N|\phi_n^{(N)}\rangle\langle\phi_n^{(N)}|.
\end{equation}
Subsequently, the total e-H wave function is expanded using
\begin{eqnarray}
|\Psi_S^{(+)}\rangle&=&(1+(-1)^SP_r)|\psi_S^{(+)}\rangle \nonumber\\&\approx&
(1+(-1)^SP_r)I_2^{(N)}|\psi_S^{(+)}\rangle,
\end{eqnarray}
where $P_r$ is the space exchange operator and $|\psi_S^{(+)}\rangle$ is an
unsymmetrized form of the total wave function. In the CCC approach
we calculate the $T$-matrix elements
\begin{eqnarray}
&&\langle k_f\phi_f^{(N)}|T_S|\phi_i^{(N)}k_i\rangle=
\langle k_f\phi_f^{(N)}|V_S|\phi_i^{(N)}k_i\rangle\nonumber\\
&&+\sum_{n=1}^N\int_0^\infty dk\frac{
\langle k_f\phi_f^{(N)}|V_S|\phi_n^{(N)}k\rangle
\langle k\phi_n^{(N)}|T_S|\phi_i^{(N)}k_i\rangle}
{E+i0-\epsilon_n^{(N)}-k^2/2},
\label{T}
\end{eqnarray}
where $V_S$ is the effective interaction potential~\cite{BS92}. Upon
solution of (\ref{T}) the ionization scattering amplitudes are defined
as~\cite{BF96} 
\begin{equation}
f_S^{(N)}(k_f,q_f)=\langle q_f^{(-)}|\phi_f^{(N)}\rangle
\langle k_f\phi_f^{(N)}|T_S|\phi_i^{(N)}k_i\rangle,
\label{amp}
\end{equation}
where $\langle q_f^{(-)}|$ is a Coulomb wave with normalization of
$\sqrt{2/\pi}$ (no $1/q_f$ factor) of energy
$q_f^2/2=\epsilon_f^{(N)}$, and such amplitudes may be defined for all 
$0<\epsilon_n^{(N)}<E$. Bencze and Chandler~\cite{BC99} are quite
happy with this definition of the ionization amplitude and claim that
in the limit of infinite $N$ it leads to the true scattering
amplitudes that must, therefore, satisfy the symmetrization relation
(\ref{symcon}). In our view their derivation has not proved either
claim as the limiting procedure ignored what happens to the
close-coupling boundary conditions. These are crucial to the
definition of the close-coupling formalism in that they allow for only 
one electron to escape to true infinity due to the $L^2$ nature of
$\phi_f^{(N)}$. Extensive comparison with
experiment~\cite{BF96,BF96l,BFRE98,B99jpbl}  has been our sole 
evidence to suggest that thus-defined amplitudes (for the full
problem) converge to the correct ionization amplitudes, but only for
$q\le k$. Let us demonstrate this by comparison with the recent
SDCS benchmark data of Baertschy {\em et al.}~\cite{BRIM99}. 

\begin{figure}
\hspace{-2truecm}\epsffile{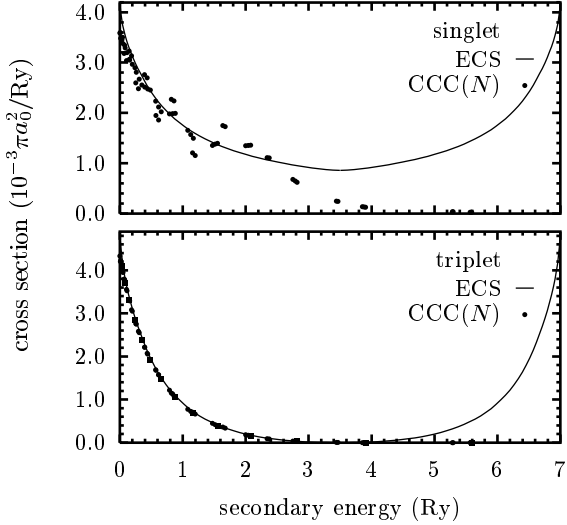}
\caption{The singlet and triplet (spin-weights included) SDCS for the
Temkin-Poet model e-H problem at a total energy $E=7$ Ry. The SDCS
calculated by the external complex scaling (ECS) method are due to
Baertschy {\em et al.}~\protect\cite{BRIM99}. The present CCC
calculations are are for $N=21,\dots,25$.
}
\label{sdcs_st}
\end{figure}
In Fig.~\ref{sdcs_st} we present the e-H model SDCS. These are
obtained from (\ref{amp}) by reference to the total ionization cross
section estimate~\cite{BF95sdcs}
\begin{eqnarray}
\sigma_{\rm ion}^{(NS)} &=& \sum_{n:0< \epsilon_n^{(N)}< E} 
|\langle k_n\phi_n^{(N)}|T_S|\phi_i^{(N)}k_i\rangle|^2\\
&=&\int_0^{\sqrt{2E}} dk |f_S^{(N)}(k,q)|^2\\
&=&\int_0^E de |f_S^{(N)}(k,q)|^2/\sqrt{2e}\\
&\equiv&\int_0^E de {\frac{d\sigma}{de}}^{(NS)} \!\!\!\!\!\!\!\!\!\!\!\!{^{\rm ion}}(e).
\label{sdcs}
\end{eqnarray}
The CCC($N$) calculations have been performed for $N=21,\dots,25$. For 
the triplet channel all of the CCC calculations lie on
the smooth curve of the ECS-calculated SDCS, but only for secondary
energies less than $E/2$. At higher energies the latter increases
symmetrically about $E/2$, whereas the CCC calculations all yield
near-zero cross sections. To our mind this is a very satisfactory
result. All of the physics is contained in the secondary energy range
$[0,E/2]$. Clearly, CCC has converged to the correct result on this
energy range, and no double counting of the ionization cross section
occurs even though the integration endpoint in (\ref{sdcs}) is $E$.

Unfortunately, the singlet case is  more complicated. The
unphysical oscillatory CCC-calculated SDCS has not converged with
increasing $N$, but oscillates about the ECS-calculated SDCS on the
$[0,E/2]$ secondary energy range. 
The ECS calculations show that the SDCS at $E/2$ is substantial,
unlike the triplet cases where it is zero due the Pauli Principle. 
We believe that as $N\to\infty$ the CCC($N$)-calculated SDCS would
converge to the step-function formed by the ECS SDCS on the secondary
energy range $[0,E/2]$ and zero elsewhere. These results are
consistent with our earlier expectations~\cite{B97l}, and with the recent
work of Miyashita, Kato, and Watanabe~\cite{MKW99}.

Let us return to the criticism of Bencze and Chandler. It is
independent of total spin, and so they argue that even our triplet
results have not converged. Somehow as $N$ is increased to infinity
they expect the CCC-calculated SDCS to converge to that calculated by
the ECS method. One consequence of this is that the unitary CCC method 
will double-count the total ionization cross section, since
(\ref{sdcs}) always holds. Another
interesting consequence is found by considering what the CCC-calculated
SDCS at large secondary energies really mean. 

For a given $E$ only pseudostates with $\epsilon_n^{(N)}<E$ may be
excited. The cross sections in Fig.~\ref{sdcs_st}, at secondary energies
near $E$, correspond to near-threshold excitation of these states. In
other words, the shape of the CCC-calculated SDCS from large to small
secondary energies shows the threshold onwards behavior of the cross
section 
\begin{equation}
\sigma_n^{(NS)}=|\langle k_n\phi_n^{(N)}|T_S|\phi_i^{(N)}k_i\rangle|^2
\end{equation}
for the excitation of the pseudostates. 

\begin{figure}
\vspace{1.0truecm}
\hspace{0truecm}\epsffile{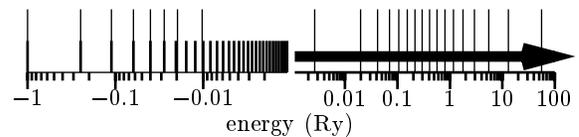}
\caption{The 25 long tics are the energy levels in the CCC(25)
calculations. The short tics and arrow indicate the true discrete
energies and the continuum, respectively.
}
\label{en25}
\end{figure}
\begin{figure}
\hspace{-1.5truecm}\epsffile{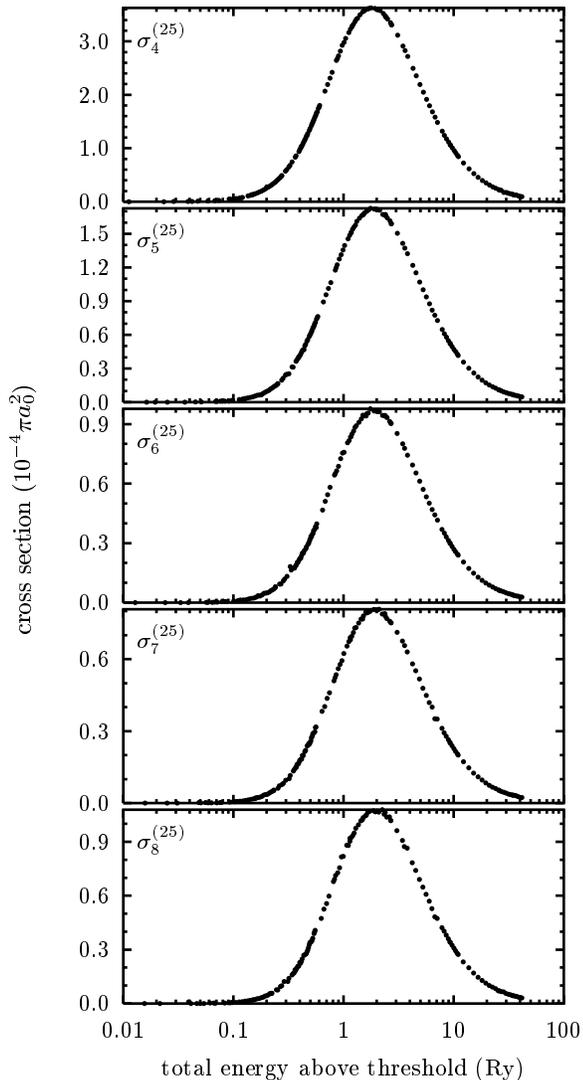}
\caption{Triplet cross sections for excitation of the specified
negative-energy states of the CCC(25) calculation in the Temkin-Poet model.
}
\label{neg}
\end{figure}

To demonstrate this
explicitly we take a single $N=25$ basis, whose energies are given in
Fig.~\ref{en25}, and perform CCC(25) calculations at very many total
energies $E$. The exponential fall-off parameter~\cite{BS92} is kept
at $\lambda=1.0$ in all cases. This yields eight negative-energy
states (lowest six good eigenstates) and 17 positive-energy
states. The cross section for the 
excitation of states with $n\le3$ has been given before~\cite{BS93l},
and so we start with $n=4$.
In Fig.~\ref{neg} we give the triplet $\sigma_n^{(NS)}$ for $4\le n\le 
8$ plotted against total energy above threshold
$E-\epsilon_n^{(25)}$. Nothing particularly remarkable is
observed. The cross sections 
start very small and rise visibly at around 0.1 Ry after
threshold. The $n=8$ cross section is of greater magnitude than the
$n=7$ owing to it attempting to take into account $n>8$ true discrete
eigenstates. 

\begin{figure}
\hspace{-1.5truecm}\epsffile{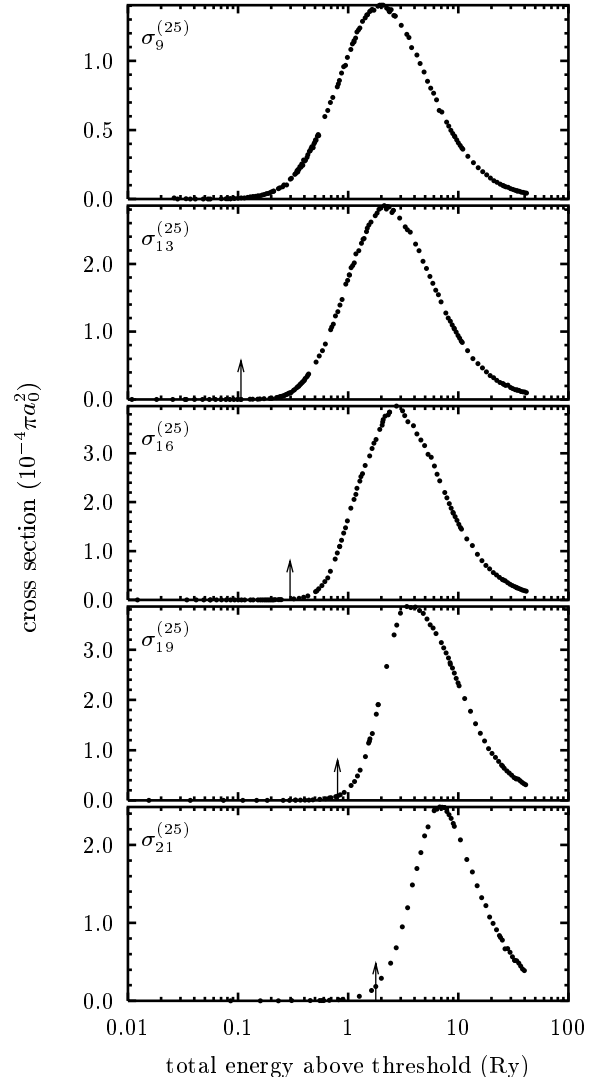}
\caption{Triplet cross sections for excitation of the specified
positive-energy states of the CCC(25) calculation in the Temkin-Poet
model. The arrows indicate the energy $\epsilon_n^{(25)}$ of the state.
}
\label{pos}
\end{figure}

What is more interesting is the behavior of the cross
sections for the excitation of positive-energy states, given in
Fig.\ref{pos}. Other than for a rise in magnitude the $n=9$ and $n=13$ 
(and those inbetween) cross
sections are much the same as the $n=8$ cross sections. This is contrary to the
expectations of Bencze and Chandler, at least in the limit of infinite 
$N$. A symmetric SDCS could only be obtained if the cross section for
exciting the positive-energy pseudostates was non-zero at threshold,
diminished to (in the present case) zero at $E/2$, and then began
to increase. Instead, as we increase
$n$ further we see that the cross section remains very small until
approximately an energy equal to $\epsilon_n^{(25)}$ above
threshold. By the Pauli Principle the true result is exactly zero at
this point. The CCC calculations give a very good approximation to
this.

Thus, we found that the bigger the energy of the pseudostate the more
energy above threshold is required before the cross section for its
excitation begins to rise. This remarkable feature demonstrates the
consistency of our interpretation of the CCC-calculated ionization
amplitudes. Whereas previously we have taken a particular $E$ and
showed, by variation of $N$, the plausibility of the step-function
hypothesis~\cite{B97l} at that $E$. Here, we have taken a single $N$
and showed, 
by variation of $E$, the plausibility of the step-function
hypothesis at all $E$, at least for the triplet channel. 

\begin{figure}
\hspace{-1.5truecm}\epsffile{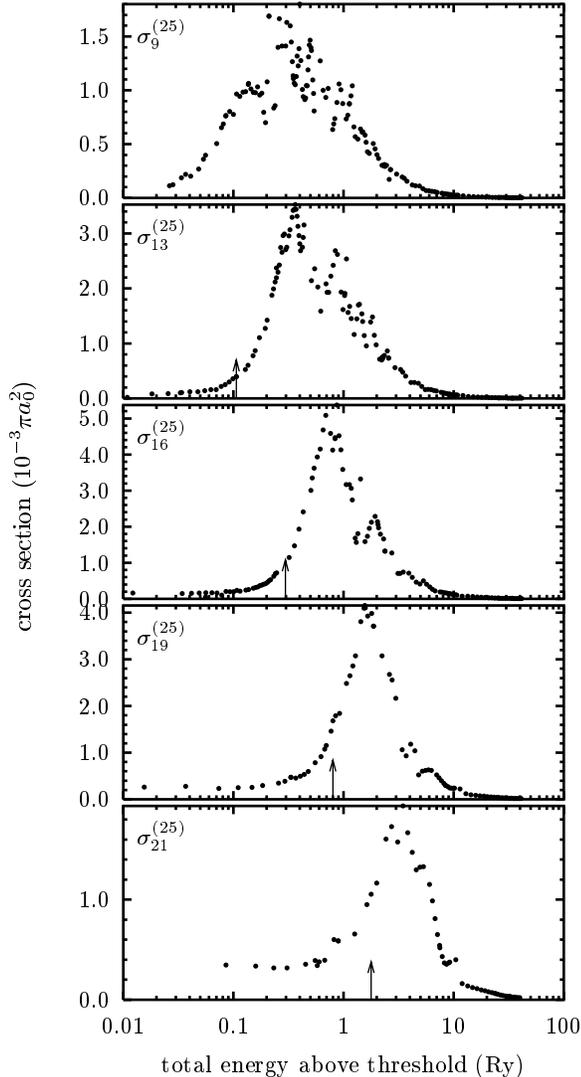}
\caption{Singlet cross sections for excitation of the specified
positive-energy states of the CCC(25) calculation in the Temkin-Poet
model. The arrows indicate  the energy $\epsilon_n^{(25)}$ of the state.
}
\label{pos0}
\end{figure}

For the
singlet channel the situation is less clear due to lack of convergence 
problems, see Fig.~\ref{pos0}. Here the cross sections rise rapidly
after the arrow (indicating $\epsilon_n^{(25)}$) and are relatively
small at the lower total energies above threshold. We suspect 
that for infinite $N$ these cross sections would be zero at energies smaller
than the energy of the state, and jump to be substantially nonzero at
higher energies.
A consequence of the present study is that we are still confident in 
the correctness of our interpretation of the application of the CCC
method to ionization processes~\cite{B97l}. We are particularly 
pleased to see the ECS theory being able to calculate accurately the
true SDCS, as these are necessary to rescale the CCC-calculated
angle-differential cross sections~\cite{B97l,B99jpb}. In the present
context the step-function in the SDCS hypothesis~\cite{B97l} may be
restated as: cross sections for the excitation of positive-energy
$\epsilon_n^{(N)}$
pseudostates remain zero past threshold ($E=\epsilon_n^{(N)}$) until
the total energy $E$ is in excess of $2\epsilon_n^{(N)}$. We suspect
that this claim is applicable to all implementations of the 
close-coupling method. 

We thank Tom Rescigno for permission to use unpublished data.
We are grateful to Emily McPherson, Ryan Coad and David Pike of the
Australian CSIRO Student Research Scheme for the many useful
discussions. Support of the Australian Research Council 
and the Flinders University of South Australia is acknowledged.
We are also indebted to the South Australian
Centre for High Performance Computing and Communications. 


\end{document}